\documentclass[aps,prd,twocolumn,groupaddress,nofootinbib,floatfix]{revtex4}

\usepackage[dvips]{graphicx}
\usepackage{amsmath,amssymb,latexsym}
\usepackage{bm}
\usepackage{epsfig}

\newcommand{\postscript}[2]{\setlength{\epsfxsize}{#2\hsize}
\centerline{\epsfbox{#1}}}

\newcommand{\md}{M_D}
\newcommand{\mbh}{M_{\text{BH}}}

\newcommand{\xmin}{x_{\text{min}}}

\newcommand{\tev}{\text{TeV}}

\def\av#1{\langle #1\rangle}
\def\neg{N_{\mu^\pm}}
\def\bn{N-\neg}
\newcommand{\Tbh}{T_{\text{BH}}}

\def\sh{\sqrt{\hat s}}

\begin{document}

\title{Black Holes at the IceCube Neutrino Telescope}

\author{Luis A. Anchordoqui}
\affiliation{Department of Physics,
University of Wisconsin-Milwaukee, P.O. Box 413, Milwaukee, WI 53201, USA
}

\author{Matthew M. Glenz}
\affiliation{Department of Physics,
University of Wisconsin-Milwaukee, P.O. Box 413, Milwaukee, WI 53201, USA
}

\author{Leonard Parker}
\affiliation{Department of Physics,
University of Wisconsin-Milwaukee, P.O. Box 413, Milwaukee, WI 53201, USA
}

\date{October 2006}
\begin{abstract}
  \noindent If the fundamental Planck scale is about a TeV and the
  cosmic neutrino flux is at the Waxman-Bahcall level, quantum black
  holes are created daily in the Antarctic ice-cap. We re-examine the
  prospects for observing such black holes with the IceCube
  neutrino-detection experiment. To this end, we first revise the black hole 
  production rate by incorporating the effects of inelasticty, i.e.,
  the energy radiated in gravitational waves by the multipole moments
  of the incoming shock waves.  After that we study in detail the
  process of Hawking evaporation accounting for the black hole's large
  momentum in the lab system. We derive the energy spectrum of the
  Planckian cloud which is swept forward with a large, ${\cal O}
  (10^6)$, Lorentz factor. (It is noteworthy that the boosted thermal
  spectrum is also relevant for the study of near-extremal supersymmetric
  black holes, which could be copiously produced at the LHC.) In the
  semiclassical regime, we estimate the average energy of the boosted
  particles to be less than 20\% the energy of the $\nu$-progenitor.
  Armed with such a constraint, we determine the discovery reach of
  IceCube by tagging on ``soft'' (relative to what one would expect
  from charged current standard model processes) muons escaping the
  electromagnetic shower bubble produced by the black hole's light
  descendants. The statistically significant $5\sigma$ excess extends
  up to a quantum gravity scale $\sim 1.3~{\rm TeV}$.
\end{abstract}


\maketitle

\section{General Idea}

Over the past few years it has become evident that a promising
approach towards reconciling the apparent mismatch of the fundamental
scales of particle physics and gravity is to modify the short distance
behavior of gravity at scales much larger than the Planck length,
$l_{\rm Pl} \sim 10^{-35}~{\rm m}$.  The key premise of such an
approach entails that the weakness of gravity is indeed evidence of
extra spatial compactified dimensions~\cite{Arkani-Hamed:1998rs}.
This is possible because standard model (SM) fields are confined to a
4-dimensional world (corresponding to our apparent universe) and only
gravity spills into the higher dimensional spacetime bulk, without
conflicting with experimental bounds~\cite{Cullen:1999hc}. Therefore,
if this new approach is correct, gravity is not intrinsically weak,
but of course appears weak at relatively large distances of common
experience because its effects are diluted by propagation in the extra
dimensions.  The distance at which the gravitational and
electromagnetic forces might have equal strength is unknown, but a
particularly interesting possibility is that it is roughly at
$10^{-19}~{\rm m}$, the distance at which electromagnetic and weak
forces are known to unify to form the electroweak force. This would
imply a fundamental D-dimensional Planck mass, $M_D \sim M_W \sim
1$~TeV, considerably smaller than the macroscopic 4-dimensional value,
$M_{\rm Pl} \sim 10^{19}~{\rm GeV}$.

If nature gracefully picked a sufficiently low-scale gravity, the
first evidence for it would likely be the observation of microscopic
black holes (BHs) produced in particle collisions~\cite{Banks:1999gd}.
According to  Thorne's hoop conjecture, a BH forms in a
two-particle collision when and only when the impact parameter is
smaller than the radius of a Schwarzschild BH of mass equal to the
total center-of-mass energy~\cite{Thorne:ji}. Subsequent to formation
a TeV-scale BH will promptly decay via thermal Hawking
radiation~\cite{Hawking:1974rv} (for $M_D = 1$~TeV, the lifetime of a
BH of mass 10 TeV is less than $10^{-25}$ s) into observable
quanta~\cite{Emparan:2000rs}.  Although the BH production cross
section, ${\cal O} (M_W^{-1})$, is about 5 orders of magnitude smaller
than QCD cross sections, ${\cal O} (\Lambda_{\rm QCD}^{-1}),$ in two
well-known papers~\cite{Dimopoulos:2001hw,Giddings:2001bu} it was
proposed that such BHs could be produced copiously at the LHC, and
that these spectacular events could be easily filtered out of the QCD
background.  This is possible by triggering on BH events with prompt
charged leptons and photons, each carrying hundreds of GeV of
energy.

Cosmic ray collisions, with center-of-mass energies ranging up to
$10^5$~GeV, certainly produce BHs if the LHC does.  The question is,
can they be detected?  Most cosmic rays are protons, which generally
collide with hadrons in the upper atmosphere, producing cascading
showers which eventually reach the Earth's surface.  At energies of
interest, however, the cosmic ray luminosity, $ L \sim
10^{-24}$~cm$^{-2}$ s$^{-1},$ is about 50 orders of magnitude smaller
than the LHC luminosity, thus making it futile to hunt for BHs in
baryonic cosmic rays.  On the other hand, neutrino interaction lengths
are still far larger than the Earth's atmospheric depth, although they
would be greatly reduced by the cross section for BH
production~\cite{Feng:2001ib}.  Cosmic neutrinos therefore would
produce BHs with roughly equal probability at any point in the
atmosphere.  As a result, the light descendants of the BH may initiate
low-altitude, quasi-horizontal showers at rates significantly higher
than SM predictions~\cite{Anchordoqui:2001ei}. Because of these
considerations the atmosphere provides a buffer against contamination
by mismeasured baryons, for which the electromagnetic channel is
filtered out.

Neutrinos that traverse the atmosphere unscathed can produce BHs
through interactions in the Antarctic ice-cap and be detected by the
IceCube neutrino telescope~\cite{Kowalski:2002gb}. This telescope,
which is currently being deployed near the Amundsen-Scott station,
comprises a cubic-kilometer of ultra-clear ice about a mile below the
South Pole surface, instrumented with long strings of sensitive photon
detectors which record light produced when neutrinos interact in the
ice~\cite{Ahrens:2002dv}. The In-ice array is complemented by IceTop,
a surface air shower detector consisting of a set of 160 frozen water
tanks, which serves as a veto for atmospheric muon background. At the
same time, the energy deposited by tagged muon bundles in air shower
cores becomes an external source of energy calibration.  Altogether,
the expected energy resolution of the experiment is about $\pm 0.1$ on
a log$_{10}$ scale.  Moreover, the energy reconstruction is optimized
for neutrino energy $E_\nu > 10^{6}$~GeV, allowing sufficient
precision ($\pm 0.2$ on a log$_{10}$ scale) to separately assign the
energy fraction for emergent muons in neutrino interactions. 
Because
of this, the inelasticity distribution of events becomes a unique tool
for SM background rejection, providing powerful discrimination of
resonant processes~\cite{Anchordoqui:2006wc}. In this work we
re-examine the prospects for discovering BH resonances at IceCube, by
tagging on ``soft'' (relative to what one would expect from charged
current SM processes) muons escaping the electromagnetic
shower bubble triggered by the BH explosion.

The paper is organized as follows. In Sec.~\ref{2} we update the
semi-classical BH production cross section considering a new
estimate~\cite{Yoshino:2005hi} of the energy radiated in gravitational
waves by the multipole moments of the incoming shock waves. This is
followed by a detailed discussion of Hawking evaporation taking into
account that cosmic neutrinos produce BHs with large momenta in the
lab system.  Specifically, we derive the energy distribution of the
Planckian cloud which is boosted in the forward direction with a large
Lorentz factor. Armed with this distribution, we estimate the average
energy of the BH light descendants to be less than 20\% the energy of
the $\nu$-progenitor.  The Hawking radiated muons then provide a very
clean signal with negligible SM background, as the production of
``soft'' leptons in charged current (CC) interactions occurs at a much
smaller rate than BH production. Our results for event rates and
discovery reach are presented in Sec.~\ref{3}. Conclusions are given
in Sec.~\ref{4}.  In this last section we also entertain the
possibility of producing TeV-scale near-extremal BHs in particle
collisions.

\section{BH Production and Evaporation}
\label{2}

Analytic and numerical studies have revealed that gravitational
collapse takes place at sufficiently high energies and small impact
parameters, as conjectured years ago by Thorne~\cite{Thorne:ji}. In
the case of 4-dimensional head-on collisions~\cite{D'Eath:hb}, as well
as those with non-zero impact parameter~\cite{Eardley:2002re}, a
horizon forms when and only when a mass is compacted into a hoop whose
circumference in every direction is less than $2 \pi$ times its
Schwarzschild radius up to a factor of order 1. In the D-dimensional
scenario the Schwarzschild radius still characterizes the maximum
impact parameter for horizon formation~\cite{Yoshino:2001ik}. In the
course of collapse, a certain amount of energy is radiated in
gravitational waves by the multipole moments of the incoming shock
waves~\cite{D'Eath:hb}, leaving a fraction $y \equiv \mbh/\sh$
available for Hawking evaporation~\cite{Hawking:1974rv}. Here, $\mbh$ is
a {\it lower bound} on the final mass of the BH and $\sh$ is the
center-of-mass energy of the colliding particles, taken as partons.
This ratio depends on the impact parameter of the collision, as well
as on the dimensionality of space-time~\cite{Yoshino:2002br}.

Of course, this work is purely in the framework of classical general
relativity, which is valid only for sufficiently massive BHs, $M_{\rm
  BH} \gg M_D$.  For masses close to $M_D$, gravity becomes strong and
the classical description can no longer be trusted. Hence, it is
important to impose a lower cutoff on the mass of microscopic BHs for
which the simple semiclassical arguments can reasonably be expected to hold.
Following~\cite{Anchordoqui:2003ug}, we define $x_{\rm min} = M_{\rm
  BH}^{\rm min}/M_D = 3,$ where $M_{\rm BH}^{\rm min}$ is the smallest
BH mass for which we trust the semiclassical approximation.

String theory provides a promising route for understanding the regime of
strong quantum gravity and in particular for computing cross sections
at energies close to the Planck scale~\cite{Dimopoulos:2001qe}.
Therefore, the ensuing discussion will be framed in the context of
string theory. To be specific we will consider an embedding of a
10-dimensional low-energy scale gravity scenario within the context of
SO(32) Type I superstring theory, where gauge and charged SM fields
can be identified with open strings localized on a 3-brane and the
gravitational sector consists of closed strings that propagate freely
in the internal dimensions of the universe~\cite{Antoniadis:1998ig}.
After compactification on $T^6$ down to four dimensions, $M_{\rm Pl}$
is related to the string scale, $M_{\rm s}$, and the string coupling
constant, $g_{\rm s}$, by $M_{\rm Pl}^2 = (2 \pi \,r_{\rm c})^6
\,M_{\rm s}^8/g_{\rm s}^2$, where $r_c$ is the compactification radius. 
Within this framework, the problem of
avoiding fast baryon decay~\cite{Adams:2000za} or lepton flavor
violation~\cite{Anchordoqui:2006xv} is shifted to the examination of
symmetries~\cite{Krauss:1988zc} in the underlying string theory which
would suppress the appropriate non-renormalizable operators at low
energies. Nevertheless, it is important to stress that for $x_{\rm
  min} \geq 3$ the typical decay involves a large number of particles.
Therefore, though these symmetries constrain the decay of the BH,
throughout this paper we ignore the constraints imposed by the few
conservation laws and we assume that BHs decay with roughly equal
probability into all SM particles. From now on we set $D = 10$.

The inclusive production of BHs proceeds through different final
states for different classical impact
parameters $b$~\cite{Yoshino:2002br}. These final states are characterized
by the fraction $y(z)$ of the initial parton center-of-mass
energy, $\sqrt{\hat s}$, which is trapped within
the horizon. Here, $z= b/b_{\rm max},$ and $b_{\rm max}=
\sqrt{F} \, r_s(\sqrt{\hat s})$ is the maximum impact parameter for collapse, 
where
\begin{equation}
\label{schwarz}
r_s(\sqrt{\hat{s}}, M_{\rm 10}) =
\frac{1}{M_{10}}
\left[ \frac{\sqrt{\hat s}}{M_{10}} \, 8 \, \pi^{3/2} \,\, \Gamma(9/2)
\right]^{1/7}
\end{equation}
is the radius of a Schwarzschild BH in 
$10$-dimensions~\cite{Myers:un}, and $F$ is a form factor.

A bound on the inelasticity and the form factor can be obtained by
studying the formation of an aparent horizon, which (because of 
cosmic censorship~\cite{Penrose:1969pc}) gurantees the formation of a
BH event horizon~\cite{Wald}. Such a study can be easily accomplished
by modeling the incoming partons as two
Aichelburg-Sexl~\cite{Aichelburg:1970dh} shock waves (i.e, by boosting
the Schwarzschild solution to the speed of light at fixed energy).
The scattering of partons is then simulated through the superposition
of two shock waves coming from opposite directions, such that their
union defines a closed trapped surface which provides a lower bound on
$M_{\rm BH}$ and $b_{\rm max}$~\cite{Yoshino:2002br}. This lower
bound, however, depends on the slice used to determine the apparent
horizon, and becomes larger if the apparent horizon is taken on the
future light cone of the collision plane~\cite{Yoshino:2005hi}. This
is because it is possible that for a given impact parameter an 
apparent horizon is not yet formed on the so-called ``old slice'', but
arises by the time a later ``new slice'' is reached. In our
calculations we consider the estimates of $y$ and $F$ in both the
``old''~\cite{Yoshino:2002br} and ``new''~\cite{Yoshino:2005hi}
slices.

The $y$ dependance complicates the parton model calculation, since the
production of a BH of mass $M_{\rm BH}$ requires that $\hat s$ be
$M_{\rm BH}^2/y^2(z)$, thus requiring the lower cutoff on parton
momentum fraction to be a function of impact parameter~\cite{note1}.
Because of the complexity of the final state, we assume that amplitude
intereference effects can be ignored and we take the $\nu N$ cross
section as an impact parameter-weighted average over parton cross
sections, with the lower parton fractional momentum cutoff determined
by $x_{\rm min}$.  This gives a lower bound ${\cal X} = (\xmin
M_{10})^2/[y^2(z)s]$ on the parton momentum fraction $x$, where
$\sqrt{s}$ is the center-of-mass energy of the $\nu N$ collision. All
in all, the $\nu N \to {\rm BH}$ cross section
reads~\cite{Anchordoqui:2003jr}
\begin{equation}
\sigma =  \int_0^1 2 z \,dz 
\int_{{\cal X}}^1 dx \, F \,  
\pi r_s^2(\sqrt{\hat{s}}, M_{\rm 10}) \,\sum_i f_i(x,Q) \ ,
\label{sigma}
\end{equation}  
where $\hat{s} = xs= 2 x m_N
E_\nu$, $i$ labels parton species, and the $f_i(x,Q)$ are parton
distribution functions (pdfs). 

\begin{figure}
 \postscript{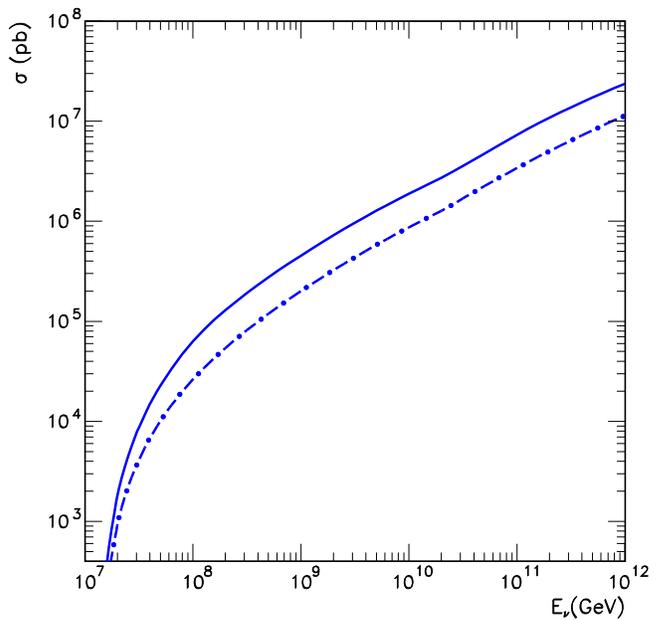}{0.98}
 \caption{BH production cross section for $x_{\rm min} = 3$ and
   $M_{10} = 1~\tev$.  The solid line indicates the result obtained
   using the new estimates of $F$ and $y$ given in~\cite{Yoshino:2005hi}, 
   whereas the dot-dashed line indicates the result
   obtained assuming the old values from Ref.~\cite{Yoshino:2002br}.}
\label{sigmaf}
\end{figure}

The choice of the momentum transfer $Q$ is governed by considering the
time or distance scale probed by the interaction. Roughly speaking,
the formation of a well-defined horizon occurs when the colliding
particles are at a distance $\sim r_s$ apart. This has led to the
advocacy of the choice $Q \simeq r_s^{-1}$~\cite{Emparan:2001kf},
which has the advantage of a sensible limit at very high energies.
However, the dual resonance picture of string
theory~\cite{Horowitz:1996nw} would suggest a choice $Q \sim
\sqrt{\hat s}$. Fortunately, as noted
elsewhere~\cite{Anchordoqui:2002vb}, the BH production cross section
is largely insensitive to the details of the choice of $Q$. In what
follows we use the CTEQ6D pdfs~\cite{Pumplin:2002vw} with $Q = {\rm
  min} \{r_s^{-1}, 10~{\rm TeV}\}$. In Fig.~\ref{sigmaf} we show the
BH production cross section for $x_{\rm min} = 3$ and $M_{10} =
1~\tev$.

BHs produced in particle collisions have non-vanishing angular momenta
determined by the impact parameter of the incoming partons.  Since $b$
is only non-zero along the brane directions, the angular momentum lies
within the brane.  Moreover, the initial horizon is likely to be very
asymmetric as only gravity spills into the compactified dimensions.
Therefore, the excited BH state carries additional hair corresponding
to multipole moments for the distribution of gauge charges and energy
momentum within the asymmetric configuration.  The decay of an excited
spinning BH state follows several stages. The initial configuration
looses hair associated with multipole moments in a balding phase
through emission of gravitational waves. In addition, gauge charges
inherited from the initial state partons are discharged within this
phase via Schwinger emission.  The subsequent spinning BH evaporates
in a two-step process: a short spin-down phase in which angular
momentum is shed~\cite{Frolov:2002xf}, followed by a long
Schwarzschild phase of semi-classical Hawking radiation.

In the rest frame of the Schwarzschild BH, both the average
number~\cite{Hawking:1974rv} and the probability distribution of the
number~\cite{Parker:1975jm} of outgoing particles in each mode obey a
thermal spectrum. In 10-dimensions, the emission rate per degree of
particle freedom $i$ of particles of spin $s$ with initial total
energy between $(\omega, \omega + d \omega)$ is found to be~\cite{Han:2002yy}
\begin{equation}
\frac{\dot{N}_i}{d \omega} = \frac{\sigma_s (\omega)
\Omega_{d-3} \omega^{d-2}}{(d-2) (2\pi)^{d-1}} \left[
e^{\omega/T} - (-1)^{2s} \right]^{-1},
\label{rate}
\end{equation}
where 
\begin{equation}
T = \frac{7}{4\,\pi\,r} 
\end{equation}
is the instantaneous Hawking temperature,
\begin{equation}
\Omega_{d-3} = \frac{2\,\pi^{(d-2)/2}}{\Gamma[(d-2)/2]}
\end{equation}
is the volume of a unit $(d-3)$-sphere, 
\begin{equation}
\label{schwarzins}
r = \frac{1}{M_{10}}
\left[ \frac{M}{M_{10}} \, 8 \, \pi^{3/2} \,\, \Gamma(9/2)
\right]^{1/7}
\end{equation}
is the instantaneous Schwarzschild radius of mass $M$,
and $\sigma_s (\omega)$ is the greybody absorption area due to the
backscattering of part of the outgoing radiation of frequency $\omega$
into the BH (a.k.a. the greybody factor)~\cite{Page:1976df}. Recall that SM
fields live on a 3-brane ($d=4$), while gravitons inhabit the entire
spacetime ($d=10$). The prevalent energies of the decay quanta are of
${\cal O}(T \sim 1/r)$, resulting in s-wave dominance of the final
state. Indeed, as the total angular momentum number of the emitted
field increases, $\sigma_s (\omega)$ rapidly gets
suppressed~\cite{Kanti:2002nr}. In the low energy limit, $\omega \, r
\ll 1,$ higher-order terms are suppressed by a factor of $3
(\omega\,r)^{-2}$ for fermions and by a factor of $25
(\omega\,r)^{-2}$ for gauge bosons. For an average particle energy
$\langle \omega \rangle$ of ${\cal O}(r^{-1})$, higher partial waves
also get suppressed, although by a smaller factor. This strongly
suggests that the BH is sensitive only to the radial coordinate and
does not make use of the extra angular modes available in the internal
space~\cite{Emparan:2000rs}. Actually, a recent detailed
analysis~\cite{Cardoso:2005vb} has explicitly shown that the relative
emission rate of SM particles and the 10-dimensional bulk graviton is
roughly 92:5. This implies that the power lost in the bulk is less
than 15\% of $M_{\rm BH}$, largely
favoring the dominance of visible decay. Therefore, in what follows we
assume the Hawking evaporation process to be dominated by SM brane modes and
we neglect graviton emission during the Schwarzschild phase. With
this in mind, the average total emission rate for particle species $i$
is,
\begin{equation}
\frac{d \av{N} }{dt}
= \frac{1}{2\pi} \left( \sum c_i\, g_i\, \Gamma_i \right)
\zeta(3)\, \Gamma(3)\, r^2\, T^3 \, ,
\label{N}
\end{equation}
where $c_i$ is the number of internal degrees of freedom of particle
species $i$, $g_i= 1\, (3/4)$ for bosons (fermions),
\begin{equation}
\Gamma_i = \frac{1}{4\pi r^2}
\int \frac{\sigma_s(\omega)\, \omega^2\, d\omega}
{e^{\omega/T}\pm 1} \left[ \int \frac{\omega^2\, d\omega}
{e^{\omega/T}\pm 1}\right]^{-1} .
\label{gamma}
\end{equation}
The rate of change of the BH mass in the evaporation process is
\begin{equation}
\left. \frac{dM}{dt} \right|_{\rm evap} = -\frac{1}{2\pi} \left(\sum c_i\, f_i\, \Phi_i\right)
 \zeta(4)\, \Gamma(4)\, r^2\, T^4\, ,
\label{M}
\end{equation}
where $f_i = 1\, (7/8)$ for bosons (fermions) and
\begin{equation}
\Phi_i = \frac{1}{4\pi r^2} \int \frac{\sigma_s(\omega)\,
\omega^3\, d\omega}
{e^{\omega/T}\pm 1} \left[ \int \frac{\omega^3\, d\omega}
{e^{\omega/T}\pm 1} \right]^{-1} .
\label{phi}
\end{equation}
Dividing Eq.~(\ref{N}) by Eq.~(\ref{M}) and integrating, one obtains a compact
expression for the average multiplicity
\begin{equation}
\av{N} = \frac{\pi}{2} 
\, \rho\, \left[ 8\, \pi^{3/2} \, \Gamma(9/2) \right]^{1/7}\, 
\left[ \frac{\mbh}{\md} \right]^{8/7}  = \rho\ S_0 \, ,
\label{multiplicity}
\end{equation}
where
\begin{equation}
\rho=\frac{\sum c_i\, g_i\, \Gamma_i}{\sum c_i\, f_i\,
\Phi_i} \frac{\zeta(3)\, \Gamma(3)} {\zeta(4)\, \Gamma(4)} \, ,
\end{equation}
and
\begin{equation}
S_0 = \frac{7}{8} \, \frac{\mbh}{\Tbh}
\end{equation}
is the initial value of the entropy in terms of the initial BH mass
and Hawking temperature $\Tbh$~\cite{Cavaglia:2003hg}.  

Before proceeding, we comment briefly on the BH rate of absorption. 
The upper limit on the cross 
section for the particle absorption is~\cite{Chamblin:2003wg}
\begin{equation}
\sigma_{\rm accr} = \pi r_{\rm eff}^2\,, 
\end{equation}
and so the accretion rate of the BH mass becomes
\begin{equation}
\left. \frac{dM}{dt}\right|_{\rm accr} =  \pi\,\, 
r^2_{\rm eff}\,\, \epsilon \,\,,
\end{equation}
where
\begin{equation}
r_{\rm eff} = \sqrt{\frac{9}{7}}\, \frac{1}{M_{10}}\, \left[36 \, \pi^{3/2}\,\, \Gamma(9/2)\, \frac{M}{M_{10}}\right]^{1/7}
\end{equation}
is the effective BH radius for capturing particles~\cite{Anchordoqui:2002cp} 
and $\epsilon$ is the nearby parton energy density. 
The net change of the BH mass is therefore
\begin{equation}
\frac{dM}{dt} = \left. \frac{dM}{dt}\right|_{\rm accr} + \left. \frac{dM}{dt}\right|_{\rm evap} \,\, .
\end{equation}
Now, using the greybody parameters given in Table~\ref{t} it is easily
seen that $dM/dt > 0 \Leftrightarrow \epsilon > 10^{10}~{\rm
  GeV/fm}^3.$ The highest earthly value of energy density of partonic
matter will be the one created at the LHC, $\epsilon_{\rm LHC} <
500~{\rm GeV/fm}^3.$ This means that the BHs that could be produced at
the South Pole would evaporate much too quickly to swallow the partons
nearby. Contrary to collider experiments, these BHs are produced
with large momentum in the lab system, and their decay products are
swept forward with large Lorentz factors. 

To perform a Lorentz transformation of the evaporating BH from
its rest frame $S$ to the rest frame $S'$ of the observer (IceCube),
we make use of the energy-momentum tensor $T_{\mu\nu}$ of the outgoing
Hawking radiation.  In the rest frame of the BH, at a distance large
with respect to the Schwarzschild radius of Eq.~(\ref{schwarzins}), the
energy-momentum tensor of the outgoing particles with energies
$\omega$ in the range $d\omega$ and with directions lying in a solid
angle $d\Omega$ is
\begin{equation}
dT^{\mu\nu}= \frac{p^{\mu} p^{\nu}}{\omega^2}
\omega \dot{N_i}
\frac{d\Omega}{4\pi} d\omega,
\label{eq:dT1}
\end{equation}
where $\dot{N_i}$ is defined by Eq.~(\ref{rate}) as the emission rate
per degree of freedom $i$ of particles having energies $\omega$ in the
range $d\omega$ and spin $s$. As noted earlier, we are working with $d
= 4$.  In spherical coordinates with the BH centered at the origin, the
4-momentum is
\begin{equation}
p^{\mu}=\omega(\sin\theta \cos\phi, \sin\theta \sin\phi, \cos\theta, 1).
\end{equation}
Integrating the (0,0) component of Eq.~(\ref{eq:dT1}) over all
directions and energies, we see that $T^{00}$ is the rate at which the
BH radiates energy per degree of freedom $i$ of a particle of spin
$s$. This requirement dictated the expression that we wrote in Eq.
(\ref{eq:dT1})~\cite{Weinberg}.

\begin{figure}
 \postscript{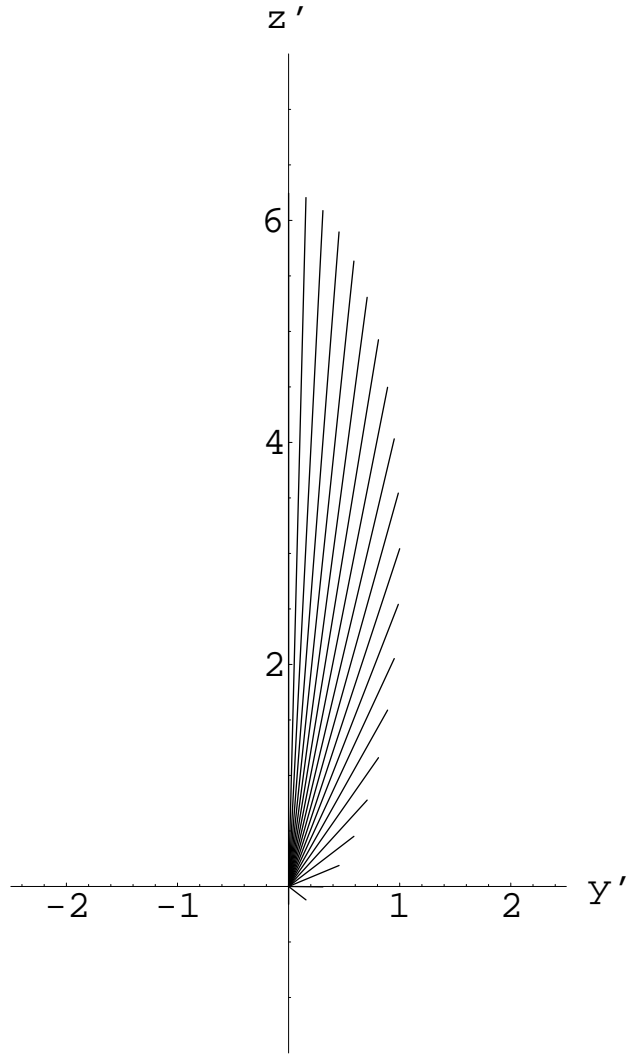}{0.98}
 \caption{Distribution of Particles emitted by a BH at the
origin and moving in the $z'$ direction at speed 0.95c. In the
BH's rest frame, the particle directions are uniformly
distributed and each particle has 1 unit of energy.
The length and angle of each line represents the energy
and angle of an emitted particle in the observer's rest
frame. Rotate the figure about the $z'$ axis for the
3-dimensional distribution.}
\label{bs}
\end{figure}

Explicitly, we have for $dT^{\mu\nu}$ in the BH's rest frame (with
$D=10$ and $d=4$):
\begin{eqnarray}
dT^{\mu\nu} & = & p^{\mu} p^{\nu}
\frac{\sigma_s(\omega)}{32\pi^3}\omega\, [\exp(\omega/T)-(-1)^{2s}]^{-1} \nonumber \\
 & \times & \sin \theta \,\, d\theta\,\, d\phi\,\, d\omega \,\, .
\end{eqnarray}
Let the black hole move at speed $v_{\rm BH}$ in the $z'$ direction
relative to the rest frame $S'$ of the observer, and take the axes of
$S$ and $S'$ to be parallel.  Then one has $\phi' = \phi$,
\begin{equation}
\omega' = \frac{\omega (1+v_{\rm BH} \cos\theta )} {(1-v_{\rm BH}^2)^{1/2}}
\label{k0}
\end{equation}
and
\begin{equation}
\cos \theta' = \frac{v_{\rm BH} + \cos\theta} {1 + v_{\rm BH} \cos\theta} \ .
\end{equation}
from which it follows that 
\begin{equation}
\sin \theta' d\theta' d\phi' = (\omega/\omega')^2 \sin \theta d\theta d\phi \ .
\end{equation}
Therefore, in the rest frame of the observer
\begin{eqnarray}
dT'^{\mu\nu} & = & p'{}^{\mu} p'{}^{\nu}
\frac{\sigma_s(\omega)}{32\pi^3} \omega' \, 
\left[\exp(\omega'/T')-(-1)^{2s}\right]^{-1} \nonumber \\
 & \times & \sin\theta'\,\, d\theta'\,\, d\phi' \,\, d\omega' \,,
\label{k1}
\end{eqnarray}
where
\begin{equation}
T' =  \frac{(1+v_{\rm BH} \cos \theta )} {(1-v_{bh}^2)^{1/2}} T
\label{k2}
\end{equation}
This can also be written using the inverse Lorentz transformation as
\begin{equation}
T' =  \frac{(1-v_{\rm BH}^2)^{1/2}} {(1-v_{\rm BH} \cos \theta' )}  T,
\end{equation}
where $\theta'$ is the angle between the direction of the black hole
and of the emitted particle as measured in the observer's rest frame.
Similarly, we can write
\begin{equation}
\omega = \frac{\omega' (1-v_{\rm BH} \cos \theta' )}{(1-v_{\rm BH}^2)^{1/2}} \, .
\end{equation}
Putting this into the grey body factor $\sigma_s(\omega)$ gives its
dependence on the energy and direction of the emitted particle as
measured in the observer's rest frame. Figure~\ref{bs} shows how particles 
tend to be emitted in the direction of motion of a moving black hole.

\begin{table}
\caption{Degrees of freedom of particle species and greybody
parameters as defined in Eqs.~(\ref{gamma}) and (\ref{phi}).}
\begin{tabular}{|c|c|c|c|}
\hline
\hline
\hspace{0.4cm} particle's spin \hspace{0.4cm} &
$\hspace{0.7cm} c_i \hspace{0.7cm}$ &
$ \hspace{0.7cm} \Gamma_i \hspace{0.7cm} $ &
$\hspace{0.7cm} \Phi_i \hspace{0.7cm} $   \\
\hline
{\small 0} & 1 & 0.80  & 0.80 \\
$\frac{1}{2}$ & 90 & 0.66 & 0.62 \\
{\small 1} & 27 & 0.60 & 0.67  \\
\hline
\hline
\end{tabular}
\label{t}
\end{table}

Despite the large Lorentz factors, it is easily seen from Eqs.~(\ref{k0}), (\ref{k1}), and (\ref{k2}) that the expected average energy of outgoing
muons from a BH explosion is much smaller than the average energy of
a secondary muon produced in a CC interaction of a neutrino
with the same energy. Specifically, the inclusive cross section for
the reaction $\nu_\mu N \to \mu^- + {\rm anything}$ is
\begin{eqnarray}
\frac{d\sigma_{\rm CC}}{dy} & = &
\frac{2\,G_{\rm F}^2 \, m_N\, E_\nu}{\pi} \,\,
\left(\frac{M_W^2}{Q^2 + M_W^2}\right)^2 \nonumber \\
 & \times & \int dx [x f_q(x, Q) + x f_{\overline q}(x, Q) (1-y)^2] \,,
\label{sigmaSM}
\end{eqnarray}
where $m_N$ is the mass of the nucleon, $x = Q^2/2m_N\nu$, $y =
\nu/E_\nu,$ with $-Q^2$ the momentum transfer between the neutrino and
muon, and $\nu$ the lepton energy loss in the lab frame, $\nu = E_\nu
- E_\mu.$ Here, $G_{\rm F} = 1.16632 \times 10^{-5}~{\rm GeV}^{-2}$ is
the Fermi constant, $M_W$ is the mass of the $W$ gauge boson, and
$f_q(x,Q)$ and $f_{\overline q} (x, Q)$ stand for combinations of quark
and anti-quark pdf's, respectively~\cite{Gandhi:1998ri}.  For
$\sqrt{s} > 5~{\rm TeV},$ the mean inelasticity is $\langle y \rangle
< 0.25$~\cite{Gandhi:1995tf}.  Therefore, SM events mimicking the
explosion of BHs with $S_0 > 10$ are far in the tail of the $y$
distributuion. Because of this, as we show in the next section, the
ratio of muon to hadronic energy deposit in a given event provides a
clean signal to search for BHs at the South Pole.

\section{IceCube Discovery Reach}
\label{3}

To evaluate the BH discovery reach at IceCube,
one has to estimate the ``beam luminosity,'' i.e. the magnitude of the
(yet to be detected!) cosmic neutrino flux. We know that cosmic accelerators
produce particles with energies in excess of $10^{11}~{\rm GeV}$ (we do
not know where or how~\cite{Anchordoqui:2002hs}), and a neutrino beam is
expected to come in association with these cosmic
rays~\cite{Gaisser:1994yf}. However, given our ignorance of the opacity of
the sources, it is difficult to calculate the magnitude of the neutrino
flux. The usual benchmark here is the so-called Waxman-Bahcall (WB) 
flux~\cite{Waxman:1998yy}
\begin{equation}
\phi_\nu  \simeq \
 6.0 \times 10^{-8}  (E_\nu/{\rm GeV})^{-2}~{\rm GeV}^{-1}\, {\rm cm}^{-2} \,
{\rm s}^{-1} \,{\rm sr}^{-1}
\end{equation}
(all flavours), which is derived assuming that neutrinos come from
transparent cosmic ray sources, that the onset of
the extragalactic component in the cosmic ray spectrum is at
$\sim 10^{10}~{\rm GeV}$, and that there is adequate transfer of energy to
pions following $pp$ collisions.  We too will use the WB flux to estimate the
event rates necessary to quantify the sensitivity to BH production. To 
evaluate the sensitivity to the assumed flux, we also consider as an upper 
limit the (AARGHW) flux~\cite{Ahlers:2005sn}, 
\begin{equation}
\phi_\nu \simeq 3.5 \times 10^{-3}\, 
(E_\nu/{\rm GeV})^{-2.54}~{\rm GeV}^{-1}\, 
{\rm cm}^{-2} \,{\rm s}^{-1} \,{\rm sr}^{-1} \,,
\end{equation}
which is expected if extragalactic cosmic rays (from transparent
sources) begin dominating the observed spectrum at energies as low as
$\sim 10^{8.6}~{\rm GeV}$, as suggested by recent 
HiRes data~\cite{Berezinsky:2002nc}. A similar enhancement in the neutrino 
flux is expected from  ``hidden'' sources which are opaque to
ultra-high energy cosmic rays~\cite{Stecker:1991vm}.

IceCube is sensitive to both downward and upward coming cosmic neutrinos.
However, to remain conservative with our statistical sample, here we
select only downward going events. To a good approximation, the expected
number of BH events at IceCube is given by
\begin{equation}
{\cal N}_{\rm BH} = 2 \pi\, n_{\rm T}\, T\, \int dE_\nu\,\,
\sigma (E_\nu)\,\, \phi_\nu (E_\nu)\ ,
\label{eventrate}
\end{equation}
where $n_{\rm T}$ is the number of target nucleons in the effective
volume and $T$ is the running time of the experiment. In our analysis
we are interested only in contained events, for which an accurate
measurement of the inelasticty can be obtained. IceCube's
effective volume for (background rejected) contained events is roughly
$1~{\rm km}^{3}$~\cite{Anchordoqui:2005is}, which corresponds to
$n_{\rm T} \simeq 5.4 \times 10^{38}.$ In order to have a good energy
resolution to determine the energy fraction in the muon track, we set
the upper limit in Eq.~(\ref{eventrate}) well below detector
saturation, $E_{\nu, {\rm max}} = 10^{10}~{\rm
  GeV}$~\cite{Halzen:2006mq}. The lower limit on this integral will be
set so as to minimize the background from SM events and consequently
depends on the infrared cutoff $x_{\rm min}$ (more on this below).  To
give an idea of the overall picture, the total number of BHs expected
to be produced within the lifetime of the experiment, for different
``beam luminosities'' and considering $E_{\nu, {\rm min}} = 10^7~{\rm
  GeV}$, are summarized in table~\ref{t2}.

\begin{table}
  \caption{Expected number of BH events for $M_{10} = 1~\tev$ and 
different values of the infrared cutoff. We have taken an integration time of 
15~yr corresponding to the lifetime of the experiment and used the new (old) 
values of $F$ and $y$. The event rates roughly scale $\propto M_{10}^{-16/7}.$}
\begin{tabular}{|c|c|c|}
\hline
\hline
\hspace{0.3cm} $x_{\rm min}$ \hspace{0.3cm} &
\hspace{0.5cm} ${\cal N}_{\rm BH}$ [WB] \hspace{0.5cm} &
 \hspace{0.5cm} ${\cal N}_{\rm BH}$ [AARGHW] \hspace{0.5cm}   \\
\hline
 3 & 43 \ \ \ \ \ (19) & 69 \ \ \ \ \ (30) \\
 4 & 34 \ \ \ \ \ (15) & 43 \ \ \ \ \ (19) \\
 5 & 27 \ \ \ \ \ (12) & 28 \ \ \ \ \ (12) \\
 6 & 22 \ \ \ \ \ \ (9) & 20 \ \ \ \ \ \ (9) \\
\hline
\hline
\end{tabular}
\label{t2}
\end{table}

In the spirit of~\cite{Dimopoulos:2001hw}, we consider the signal of
BH events with total multiplicity $N\ge 4$ and at least one
$\mu^{\pm}$ in the final state.  To implement the first cut we make
use of the average multiplicities $\av{N}$ for the various particle
species (incorporating evolution effects during Hawking radiation)
summarized in the previous section. To implement the second cut we
define the average multiplicity for any subset of states $\{s\}$
as usual, $\langle N_{\{s\}}\rangle = B_{\{s\}} \av{N}$, where
\begin{equation}
B_{\{s\}} = \frac{\sum_{i \in \{s\}} c_i\, g_i\, \Gamma_i}
{\sum_i c_i\, g_i\, \Gamma_i} \, 
\label{branching}
\end{equation}
is the so-called ``branching fraction''.  Now, using the the parameters 
given in Table~\ref{t}, we find
$\av{N}=0.30\, M/T$ and $\av{\neg} = 0.022 \, \av{N} = 0.007 \, M/T$.

$\av{N}$ is the average value of a Poisson distribution.  If all
species are Poisson distributed, then the sum of particles in any
subset is also Poisson distributed, and so $N$, $\neg$, and $N-\neg$
are all Poisson distributed, where $\neg$ is the total number of
$\mu^{\pm}$ per event.  The signal probability (i.e., 
that a given event has $\neg \ge 1$ and $N \ge 4$) is~\cite{Anchordoqui:2003ug}
\begin{eqnarray}
P_{\rm sig} & = &  -  e^{-\av{\neg}} \left(1 - e^{-\av{\bn}}
\sum_{i=0}^3\frac{\av{\bn} ^i}{i!}\right) \nonumber\\
              & + 
& \left(1- e^{-\av{N}} \sum_{i=0}^{3}\frac{\av{N} ^i}{i\, !}\right) \,\, ,
\end{eqnarray}
and so the number of signal events becomes
\begin{equation}
{\cal N}_{\rm sig} = 2 \pi\, n_{\rm T}\, T\, \int dE_\nu\,\,
\sigma (E_\nu)\,\, \phi_\nu (E_\nu)\ P_{\rm sig}\,  .
\label{signal}
\end{equation}
The quarks and gluons emitted by the BH promptly fragment into hadrons
(mostly $\pi^\pm$ and $\pi^0$). For $E_\pi > 1~\tev$, the interaction
mean free path of $\pi^\pm$ in ice is orders of magnitude smaller than
the pion decay length, and so nearly all the hadronic energy is
channeled into electromagnetic modes through $\pi^0$ decay.  The
signal of such a hadronic/electromagnetic cascade is a bright, point
like, source of \v Cerenkov light. The shower topology can be easily
identified by the sphericity of the light pattern. The measurement of
the radius of the lightpool mapped by the lattice of photomultiplier
tubes determines the energy and turns IceCube into a total absorption
calorimeter~\cite{note}. On the other hand, the muons emitted by the
BH would produce a track moving outwards from the interaction vertex,
providing very useful tags for the event.

As discussed in the previous section the SM background masking BH
events are in the tail of the CC $y$ distribution: for $\langle N \rangle
> 4,$ the average energy of the emitted muon (after considering energy
losses due to classical radiation) is less than
20\% of the incoming neutrino energy. Therefore, to filter the background 
we evaluate Eq.~(\ref{sigmaSM}) for $y>0.8$, yielding 
\begin{equation}
\sigma_{\rm CC}^{y>0.8} \simeq 1.2 \ \ (E_\nu/{\rm GeV})^{0.358}~{\rm pb} 
\,\, .
\end{equation}
Now, substituting the CC SM cross section (with $y>0.8$) into
Eq.~(\ref{eventrate}) leads to a straightforward calculation that
shows that for $E_{\nu, {\rm min}} = 10^8~{\rm GeV}$ the expected
number of background events is negligible, less than 1 event in 15
years, independently of the selected (WB/AARGHW) beam
luminosity~\cite{Learned:1994wg}. The sensitivity of IceCube to probe
the $x_{\rm min}/M_{10}$ parameter space at the $3\sigma$
level~\cite{Feldman:1997qc} is summarized in Table~\ref{t3}.

\begin{table}
  \caption{Sensitivity of IceCube at the 3$\sigma$ level for the value of 
    $M_{10}/{\rm TeV}$ using fiducial beam luminosities. We have taken an 
    integration time of 
    15~yr corresponding to the lifetime of the 
    experiment and used the new (old) values of $F$ and $y$.}
\begin{tabular}{|c|c|c|}
\hline
\hline
\hspace{0.3cm} $x_{\rm min}$ \hspace{0.3cm} &
\hspace{0.2cm} $M_{10}/$TeV [WB] \hspace{0.2cm} &
 \hspace{0.2cm} $M_{10}/$TeV [AARGHW] \hspace{0.2cm}   \\
\hline
 3 & 1.5 \ \ \ \ \ (1.2) & 1.5 \ \ \ \ \  (1.2) \\
 5 & 1.3 \ \ \ \ \ (1.1) & 1.3 \ \ \ \ \  (1.1) \\
 7 & 1.2 \ \ \ \ \ (1.0) & 1.2 \ \ \ \ \  (1.0) \\
 9 & 1.1 \ \ \ \ \ (1.0) & 1.1 \ \ \ \ \  (0.9) \\
\hline
\hline
\end{tabular}
\label{t3}
\end{table}

\begin{figure}
 \postscript{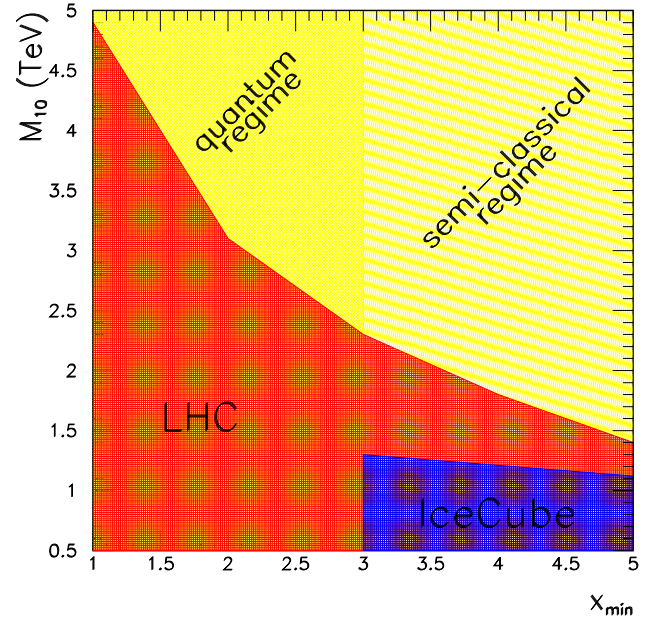}{0.98}
 \caption{IceCube discovery reach of quantum BHs, assuming the new
   estimates of $F$ and $y$ given in~\cite{Yoshino:2005hi}. For
   comparison the LHC discovery reach, assuming a cumulative
   integrated luminosity of 1~ab$^{-1}$ over the life of the collider,
   is also shown. The LHC discovery reach has been obtained by scaling
   up the results given in~\cite{Anchordoqui:2003ug}, to account for
   the new values of $y$ and $F$.}
\label{reach}
\end{figure}

The expected number of background events rises with decreasing the
low-energy cutoff. For example, for $E_{\nu, {\rm min}} = 10^7~{\rm
  GeV}$ and a beam luminosity at the AARGHW level, ${\cal N}_{\rm
  bk'd} = 10.$ To remain conservative, we adopt this energy range and
require a $5\sigma$ excess for discovery. The resulting reach is shown
in Fig.~\ref{reach}.  

The BH entropy is a measure of the validity of the semi-classical
approximation. For $x_{\rm min} >3,$ $S_0 \gg 10$ yields small thermal
fluctuations in the emission process~\cite{Preskill:1991tb}.
Therefore, strong quantum gravity effects may be safely neglected in
this ``energy regime,'' $M_{\rm BH}^{\rm min}/M_{10}>3$. Moreover,
gravitational effects due to brane back-reaction are expected to be
insignificant for $M_{\rm BH}$ well beyond the brane tension, which is
presumably on the order of $M_{10}$~\cite{Anchordoqui:2003ug}. As
noted in the previous section, string theory provides a more complete
picture for $M_{\rm BH}$ close to $M_{10}.$ In string theory, the
ultimate fate of the BH is determined by the string
$\rightleftharpoons$ BH correspondance principle: when the
Schwarzschild radius of the BH shrinks to the fundamental string
length $l_s \gg 10^{-19}~{\rm m}$ an adiabatic transition occurs to a
massive superstring mode~\cite{Horowitz:1996nw}. Subsequent energy
loss continues as thermal radiation at the unchanging Hagedorn
temperature~\cite{Amati:1999fv}. The continuity of the cross section
at the correspondance point, parametrically in both the energy and the
string coupling, provides independent support for this
picture~\cite{Dimopoulos:2001qe}. In the perturbative string regime,
however, the parton-parton cross section contains the Chan Paton
factors which control the projection of the initial state onto the
string spectrum.  In general, this projection is not uniquely
determined by the low-lying particle spectrum, yielding one or more
arbitrary constants~\cite{Cornet:2001gy}.  Interestingly, the
parton-parton cross section derived in~\cite{Dimopoulos:2001qe} from
the Virasoro-Shapiro amplitude leads to an enhancement of the
predicted BH cross section for $1<x_{\rm min}
<3$~\cite{Anchordoqui:2003ur}. This makes it plausible to adopt the BH
cross section as a lower bound of the real Planckian cross section.
However, it is important to stress that the proposed signal to search
for BHs at IceCube strongly depends on the large density of quantum
mechanical states of BHs with $S_0 \gg 10$, and consequently it is
only valid in the semi-classical regime.  The LHC signal (hard photons
and leptons in the collision rest frame), however, can be mantained
under plausible hypotheses on the superstring decay modes. With this
in mind, the LHC discovery reach for the quantum regime shown in
Fig.~\ref{reach} may be taken as a lower bound, derived on the
assumption that the BH cross section provides a lower limit on the
string cross section.

\section{Conclusions}
\label{4}

In this work we have reviewed the possibility of searching for BHs
using cosmic neutrino interactions in the Antarctic ice. We have shown
that the ability of IceCube to accurately measure the inelasticity
distribution of events provides a unique discriminator for SM
background rejection~\cite{notetau}, allowing extremely sensitive
probes of TeV-scale BH production.  In the optimistic case that the
neutrino flux is at the WB level, IceCube has a substantial discovery
potential for BHs, well in the semi-classical regime where $M_{\rm
  BH}^{\rm min} > 3 M_{10}$. The statistically significant $5\sigma$
excess extends up to $M_{10} \sim 1.3~{\rm TeV}.$

A point worth noting at this juncture: In assessing the discovery
potential of IceCube we have performed a Lorentz transformation of the
evaporating BH from its rest frame to the lab frame, so as to obtain
the boosted energy spectrum of the outgoing Hawking radiation. Such a
spectrum is also relevant for the study of near-extremal
supersymmetric BHs, which could be copiously produced at the LHC.
These BHs can be associated with the supermassive string modes
expected to populate the quantum regime~\cite{Duff:1994jr}. Note that
in order to mantain the configuration of the initial state with ``zero
supersymmetry,'' the central charge conservation would force these BHs
to be pair produced with non-zero transverse momentum ($p_T$),
travelling along the beam pipe. Therefore, the prompt decay of the BHs
would produce a startlingly clean signal that should have very few
backgrounds: their decay products (that of course may include
sparticles~\cite{Chamblin:2004zg}), would trigger high multiplicity
fireworks with boosted spherical shape collimated into back-to-back
pencil beams. A crude estimate of the event rates can be obtained from
the analysis of ``$ij \to {\rm BH} + {\rm others}$''
subprocesses~\cite{Cheung:2001ue}. The results are encouraging: the
production rate of BHs with large $p_T > 500~{\rm GeV}$ (which is
balanced by the momentum of an energetic parton) would still be large
enough for detection at LHC (assuming an integrated luminosity
100~fb$^{-1}$). Of course the energy requirement for BH pair
production would yield an additional suppression factor and its proper
consideration requires a Monte Carlo simulation. Therefore, we
strongly recommend to include in future versions of the BH event
generators {\sc charybdis}~\cite{Harris:2003db} and {\sc
  catfish}~\cite{Cavaglia:2006uk} a detailed treatment of production and
evaporation of near-extremal supersymmetric BHs.

In closing, it is important to stress that this analysis is meant to
investigate the underlying principles and does not account for the
details of the detector response. Though the devil is generally in the
details, we believe our estimate of the IceCube discovery reach is
conservative, as we have not included the $\tau$-channel which has the
potential of nearly doubling our signal event background.
Specifically, $\tau$ leptons emitted by the BH may decay in flight
inside the instrumented volume after escaping the electromagnetic
shower bubble, thereby triggering a ``second
bang~\cite{Cardoso:2004zi}.''  The ratio of the first to second bang
fractional energy provides a clean direct signal of BH production, and
like the muon channel is ``independent'' of the absolute neutrino
flux. The inclusion of the tau channel, however, requires a full blown
Monte Carlo simulation to properly determine the
acceptance for such events, in which the triggering probability for
the second bang depends on $T_{\rm BH}$ and the corresponding tau
decay length.

In summary, over the next few years high-statistics high-energy
precision data will be collected at the LHC. In addition, the IceCube
neutrino telescope is coming on line with complementary information at
ultra-high energies. This new arsenal of data will certainly provide
an ideal testing ground for TeV-scale BH production, and, at the same
time, a unique opportunity to view similar physics from two different
points of view. Should the LHC find evidence of BHs a bit outside the
range accessible for the baseline IceCube design, the ideas discussed
in this paper could constitute another compelling reason for pursuing
HyperCube~\cite{Halzen:2003fi}.

\acknowledgments{\noindent We would like to thank Carlos Nu\~nez for
  valuable discussions on near-extremal black holes. LAA is thankful
  to Jonathan Feng, Haim Goldberg, and Al Shapere for enlightening
  discussions on black holes. The research of LAA and LP has been
  partially supported by the US National Science Foundation (NSF) and
  the University of Wisconsin-Milwaukee through the Research Growth
  Initiative (RGI) Program.  The research of MMG is supported by a
  Bradley Fellowship from the Lynde and Harry Bradley Foundation.}


\begin{thebibliography}{99}


\bibitem{Arkani-Hamed:1998rs}
  N.~Arkani-Hamed, S.~Dimopoulos and G.~R.~Dvali,
  Phys.\ Lett.\ B {\bf 429}, 263 (1998)
  [arXiv:hep-ph/9803315];
  L.~Randall and R.~Sundrum,
  Phys.\ Rev.\ Lett.\  {\bf 83}, 3370 (1999)
  [arXiv:hep-ph/9905221].


\bibitem{Cullen:1999hc} See e.g.,
  S.~Cullen and M.~Perelstein,
  Phys.\ Rev.\ Lett.\  {\bf 83}, 268 (1999)
  [arXiv:hep-ph/9903422];
  B.~Abbott {\it et al.}  [D0 Collaboration],
  Phys.\ Rev.\ Lett.\  {\bf 86}, 1156 (2001)
  [arXiv:hep-ex/0008065];
  C.~D.~Hoyle, U.~Schmidt, B.~R.~Heckel, E.~G.~Adelberger, J.~H.~Gundlach, 
  D.~J.~Kapner and H.~E.~Swanson,
  Phys.\ Rev.\ Lett.\  {\bf 86}, 1418 (2001)
  [arXiv:hep-ph/0011014];
  L.~A.~Anchordoqui, J.~L.~Feng, H.~Goldberg and A.~D.~Shapere,
  Phys.\ Rev.\ D {\bf 65}, 124027 (2002)
  [arXiv:hep-ph/0112247];
  V.~M.~Abazov {\it et al.}  [D0 Collaboration],
  Phys.\ Rev.\ Lett.\  {\bf 90}, 251802 (2003)
  [arXiv:hep-ex/0302014];
  S.~Hannestad and G.~G.~Raffelt,
  Phys.\ Rev.\ D {\bf 67}, 125008 (2003)
  [Erratum-ibid.\ D {\bf 69}, 029901 (2004)]
  [arXiv:hep-ph/0304029];
  E.~J.~Ahn, M.~Cavaglia and A.~V.~Olinto,
  Astropart.\ Phys.\  {\bf 22}, 377 (2005)
  [arXiv:hep-ph/0312249];
  V.~M.~Abazov {\it et al.}  [D0 Collaboration],
  Phys.\ Rev.\ Lett.\  {\bf 95}, 161602 (2005)
  [arXiv:hep-ex/0506063];
  S.~Hussain and D.~W.~McKay,
  Phys.\ Lett.\ B {\bf 634}, 130 (2006)
  [arXiv:hep-ph/0510083].




\bibitem{Banks:1999gd}
  T.~Banks and W.~Fischler,
  arXiv:hep-th/9906038.
  For a comprehensive review see e.g., 
  G.~Landsberg,
  J.\ Phys.\ G {\bf 32}, R337 (2006)
  [arXiv:hep-ph/0607297];
  P.~Kanti,
  Int.\ J.\ Mod.\ Phys.\ A {\bf 19}, 4899 (2004)
  [arXiv:hep-ph/0402168];
  M.~Cavaglia,
  Int.\ J.\ Mod.\ Phys.\ A {\bf 18}, 1843 (2003)
  [arXiv:hep-ph/0210296].





\bibitem{Thorne:ji}
K.~S.~Thorne,
in {\it Magic Without Magic: John Archibald Wheeler}, edited by J. Klauder
(Freeman, San Francisco, 1972) p.231.




\bibitem{Emparan:2000rs}
  R.~Emparan, G.~T.~Horowitz and R.~C.~Myers,
  Phys.\ Rev.\ Lett.\  {\bf 85}, 499 (2000)
  [arXiv:hep-th/0003118].



\bibitem{Dimopoulos:2001hw}
  S.~Dimopoulos and G.~Landsberg,
  Phys.\ Rev.\ Lett.\  {\bf 87}, 161602 (2001)
  [arXiv:hep-ph/0106295].

\bibitem{Giddings:2001bu}
  S.~B.~Giddings and S.~D.~Thomas,
  Phys.\ Rev.\ D {\bf 65}, 056010 (2002)
  [arXiv:hep-ph/0106219].





 \bibitem{Feng:2001ib}
  J.~L.~Feng and A.~D.~Shapere,
  Phys.\ Rev.\ Lett.\  {\bf 88}, 021303 (2002)
  [arXiv:hep-ph/0109106].




\bibitem{Anchordoqui:2001ei}
  L.~Anchordoqui and H.~Goldberg,
  Phys.\ Rev.\ D {\bf 65}, 047502 (2002)
  [arXiv:hep-ph/0109242];
  A.~Ringwald and H.~Tu,
  Phys.\ Lett.\ B {\bf 525}, 135 (2002)
  [arXiv:hep-ph/0111042];
  S.~I.~Dutta, M.~H.~Reno and I.~Sarcevic,
  Phys.\ Rev.\ D {\bf 66}, 033002 (2002)
  [arXiv:hep-ph/0204218];
  E.~J.~Ahn, M.~Ave, M.~Cavaglia and A.~V.~Olinto,
  Phys.\ Rev.\ D {\bf 68}, 043004 (2003)
  [arXiv:hep-ph/0306008];
  A.~Cafarella, C.~Coriano and T.~N.~Tomaras,
  JHEP {\bf 0506}, 065 (2005)
  [arXiv:hep-ph/0410358];
  L.~Anchordoqui, T.~Han, D.~Hooper and S.~Sarkar,
  Astropart.\ Phys.\  {\bf 25}, 14 (2006)
  [arXiv:hep-ph/0508312].


\bibitem{Kowalski:2002gb}
  M.~Kowalski, A.~Ringwald and H.~Tu,
  Phys.\ Lett.\ B {\bf 529}, 1 (2002)
  [arXiv:hep-ph/0201139];
  J.~Alvarez-Muniz, J.~L.~Feng, F.~Halzen, T.~Han and D.~Hooper,
  Phys.\ Rev.\ D {\bf 65}, 124015 (2002)
  [arXiv:hep-ph/0202081].


\bibitem{Ahrens:2002dv}
  J.~Ahrens {\it et al.}  [The IceCube Collaboration],
  Nucl.\ Phys.\ Proc.\ Suppl.\  {\bf 118}, 388 (2003)
  [arXiv:astro-ph/0209556].


\bibitem{Anchordoqui:2006wc}
  L.~A.~Anchordoqui, C.~A.~G.~Canal, H.~Goldberg, D.~G.~Dumm and F.~Halzen,
  arXiv:hep-ph/0609214.



\bibitem{Yoshino:2005hi}
  H.~Yoshino and V.~S.~Rychkov,
  Phys.\ Rev.\ D {\bf 71}, 104028 (2005)
  [arXiv:hep-th/0503171].









\bibitem{D'Eath:hb} R.~Penrose, unpublished (1974);
P.~D.~D'Eath and P.~N.~Payne,
Phys.\ Rev.\ D {\bf 46}, 658 (1992);
P.~D.~D'Eath and P.~N.~Payne,
Phys.\ Rev.\ D {\bf 46}, 675 (1992);
P.~D.~D'Eath and P.~N.~Payne,
Phys.\ Rev.\ D {\bf 46}, 694 (1992).

\bibitem{Eardley:2002re}
D.~M.~Eardley and S.~B.~Giddings,
Phys.\ Rev.\ D {\bf 66}, 044011 (2002)
[arXiv:gr-qc/0201034].

\bibitem{Yoshino:2001ik}
H.~Yoshino, Y.~Nambu and A.~Tomimatsu,
Phys.\ Rev.\ D {\bf 65}, 064034 (2002)
[arXiv:gr-qc/0109016].

\bibitem{Hawking:1974rv}
  S.~W.~Hawking,
  Nature {\bf 248}, 30 (1974);
  S.~W.~Hawking,
  Commun.\ Math.\ Phys.\  {\bf 43}, 199 (1975)
  [Erratum-ibid.\  {\bf 46}, 206 (1976)].


\bibitem{Yoshino:2002br}
H.~Yoshino and Y.~Nambu,
Phys.\ Rev.\ D {\bf 66}, 065004 (2002)
[arXiv:gr-qc/0204060];
H.~Yoshino and Y.~Nambu,
Phys.\ Rev.\ D {\bf 67}, 024009 (2003)
[arXiv:gr-qc/0209003].





\bibitem{Anchordoqui:2003ug}
  L.~A.~Anchordoqui, J.~L.~Feng, H.~Goldberg and A.~D.~Shapere,
  Phys.\ Lett.\ B {\bf 594}, 363 (2004)
  [arXiv:hep-ph/0311365].



\bibitem{Dimopoulos:2001qe}
  S.~Dimopoulos and R.~Emparan,
  Phys.\ Lett.\ B {\bf 526}, 393 (2002)
  [arXiv:hep-ph/0108060].

\bibitem{Antoniadis:1998ig}
  I.~Antoniadis, N.~Arkani-Hamed, S.~Dimopoulos and G.~R.~Dvali,
  Phys.\ Lett.\ B {\bf 436}, 257 (1998)
  [arXiv:hep-ph/9804398].




\bibitem{Adams:2000za}
  F.~C.~Adams, G.~L.~Kane, M.~Mbonye and M.~J.~Perry,
  Int.\ J.\ Mod.\ Phys.\ A {\bf 16}, 2399 (2001)
  [arXiv:hep-ph/0009154].

\bibitem{Anchordoqui:2006xv}
  L.~A.~Anchordoqui,
  arXiv:hep-ph/0610025.


\bibitem{Krauss:1988zc}
  L.~M.~Krauss and F.~Wilczek,
  Phys.\ Rev.\ Lett.\  {\bf 62}, 1221 (1989);
  L.~E.~Ibanez and G.~G.~Ross,
  Nucl.\ Phys.\ B {\bf 368}, 3 (1992).
  S.~W.~Hawking,
  Phys.\ Rev.\ D {\bf 72}, 084013 (2005)
  [arXiv:hep-th/0507171].






\bibitem{Myers:un}
R.~C.~Myers and M.~J.~Perry,
Annals Phys.\  {\bf 172}, 304 (1986);
P.~C.~Argyres, S.~Dimopoulos and J.~March-Russell,
Phys.\ Lett.\ B {\bf 441}, 96 (1998).
[arXiv:hep-th/9808138]




\bibitem{Penrose:1969pc}
  R.~Penrose,
  Riv.\ Nuovo Cim.\  {\bf 1}, 252 (1969)
  [Gen.\ Rel.\ Grav.\  {\bf 34}, 1141 (2002)].


\bibitem{Wald} R. Wald, {\em General Relativity,} (University of
  Chicago Press, Chicago, 1984) p.310.



\bibitem{Aichelburg:1970dh}
  P.~C.~Aichelburg and R.~U.~Sexl,
  Gen.\ Rel.\ Grav.\  {\bf 2}, 303 (1971).








\bibitem{note1} The validity of an impact parameter description in the
semi-classical calculation can be ascertained by calculating the
maximum angular momentum $J_{\rm max}=(\sqrt{\hat s}/2)b_{\rm max}$ in
a given collision: for $M_{\rm BH}\ge M_D, $ one finds $J_{\rm max}\ge
7 \gg 1$.


\bibitem{Anchordoqui:2003jr}
  L.~A.~Anchordoqui, J.~L.~Feng, H.~Goldberg and A.~D.~Shapere,
  Phys.\ Rev.\ D {\bf 68}, 104025 (2003)
  [arXiv:hep-ph/0307228].

\bibitem{Emparan:2001kf}
  R.~Emparan, M.~Masip and R.~Rattazzi,
  Phys.\ Rev.\ D {\bf 65}, 064023 (2002)
  [arXiv:hep-ph/0109287].


\bibitem{Horowitz:1996nw}
G.~T.~Horowitz and J.~Polchinski,
Phys.\ Rev.\ D {\bf 55}, 6189 (1997)
[arXiv:hep-th/9612146];
T.~Damour and G.~Veneziano,
Nucl.\ Phys.\ B {\bf 568}, 93 (2000)
[arXiv:hep-th/9907030].



\bibitem{Anchordoqui:2002vb}
  L.~A.~Anchordoqui, J.~L.~Feng, H.~Goldberg and A.~D.~Shapere,
  Phys.\ Rev.\ D {\bf 66}, 103002 (2002)
  [arXiv:hep-ph/0207139].


\bibitem{Pumplin:2002vw}
J.~Pumplin, D.~R.~Stump, J.~Huston, H.~L.~Lai, P.~Nadolsky and W.~K.~Tung,
JHEP {\bf 0207}, 012 (2002)
[arXiv:hep-ph/0201195];
  D.~Stump, J.~Huston, J.~Pumplin, W.~K.~Tung, H.~L.~Lai, S.~Kuhlmann 
  and J.~F.~Owens,
  JHEP {\bf 0310}, 046 (2003)
  [arXiv:hep-ph/0303013].





\bibitem{Frolov:2002xf}
V.~P.~Frolov and D.~Stojkovic,
Phys.\ Rev.\ D {\bf 67}, 084004 (2003)
[arXiv:gr-qc/0211055];
Phys.\ Rev.\ D {\bf 68}, 064011 (2003)
V.~P.~Frolov, D.~V.~Fursaev and D.~Stojkovic,
JHEP {\bf 0406}, 057 (2004)
[arXiv:gr-qc/0403002];
V.~P.~Frolov, D.~V.~Fursaev and D.~Stojkovic,
Class.\ Quant.\ Grav.\  {\bf 21}, 3483 (2004)
[arXiv:gr-qc/0403054];



\bibitem{Parker:1975jm}
  L.~Parker,
  Phys.\ Rev.\ D {\bf 12}, 1519 (1975);
  R.~M.~Wald,
  Commun.\ Math.\ Phys.\  {\bf 45}, 9 (1975);
  S.~W.~Hawking,
  Phys.\ Rev.\ D {\bf 14}, 2460 (1976).



\bibitem{Han:2002yy}
T.~Han, G.~D.~Kribs and B.~McElrath,
Phys.\ Rev.\ Lett.\  {\bf 90}, 031601 (2003)
[arXiv:hep-ph/0207003].


\bibitem{Page:1976df}
  D.~N.~Page,
  Phys.\ Rev.\ D {\bf 13}, 198 (1976).


\bibitem{Kanti:2002nr}
P.~Kanti and J.~March-Russell,
Phys.\ Rev.\ D {\bf 66}, 024023 (2002)
[arXiv:hep-ph/0203223];
P.~Kanti and J.~March-Russell,
Phys.\ Rev.\ D {\bf 67}, 104019 (2003)
[arXiv:hep-ph/0212199];
C.~M.~Harris and P.~Kanti,
JHEP {\bf 0310}, 014 (2003)
[arXiv:hep-ph/0309054].







\bibitem{Cardoso:2005vb}
  V.~Cardoso, M.~Cavaglia and L.~Gualtieri,
  Phys.\ Rev.\ Lett.\  {\bf 96}, 071301 (2006)
  [Erratum-ibid.\  {\bf 96}, 219902 (2006)]
  [arXiv:hep-th/0512002];
  V.~Cardoso, M.~Cavaglia and L.~Gualtieri,
  JHEP {\bf 0602}, 021 (2006)
  [arXiv:hep-th/0512116].

\bibitem{Cavaglia:2003hg} This dynamycal treatment of Hawking evaporation
introduced in Ref.~\cite{Anchordoqui:2003ug}, extends the discussion 
by M.~Cavaglia,
Phys.\ Lett.\ B {\bf 569}, 7 (2003)
[hep-ph/0305256], 
incorporating the correct energy dependence of $\sigma_s.$


\bibitem{Chamblin:2003wg}
  A.~Chamblin, F.~Cooper and G.~C.~Nayak,
  Phys.\ Rev.\ D {\bf 69}, 065010 (2004)
  [arXiv:hep-ph/0301239].


\bibitem{Anchordoqui:2002cp} 
  L.~Anchordoqui and H.~Goldberg,
  Phys.\ Rev.\ D {\bf 67}, 064010 (2003)
  [arXiv:hep-ph/0209337].


\bibitem{Weinberg} S. Weinberg, {\em Gravitation and Cosmology}, (John
  Wiley \& Sons, 1972) p.522.





\bibitem{Gandhi:1998ri}
  R.~Gandhi, C.~Quigg, M.~H.~Reno and I.~Sarcevic,
  Phys.\ Rev.\ D {\bf 58}, 093009 (1998)
  [arXiv:hep-ph/9807264];
  L.~A.~Anchordoqui, A.~M.~Cooper-Sarkar, D.~Hooper and S.~Sarkar,
  Phys.\ Rev.\ D {\bf 74}, 043008 (2006)
  [arXiv:hep-ph/0605086].


\bibitem{Gandhi:1995tf}
  R.~Gandhi, C.~Quigg, M.~H.~Reno and I.~Sarcevic,
  Astropart.\ Phys.\  {\bf 5}, 81 (1996)
  [arXiv:hep-ph/9512364].


\bibitem{Anchordoqui:2002hs}
  L.~Anchordoqui, T.~Paul, S.~Reucroft and J.~Swain,
  Int.\ J.\ Mod.\ Phys.\ A {\bf 18}, 2229 (2003)
  [arXiv:hep-ph/0206072].

\bibitem{Gaisser:1994yf}
  T.~K.~Gaisser, F.~Halzen and T.~Stanev,
  Phys.\ Rept.\  {\bf 258}, 173 (1995)
  [Erratum-ibid.\  {\bf 271}, 355 (1996)]
  [arXiv:hep-ph/9410384];
  J.~G.~Learned and K.~Mannheim,
  Ann.\ Rev.\ Nucl.\ Part.\ Sci.\  {\bf 50}, 679 (2000);
  F.~Halzen and D.\ Hooper,
  Rept.\ Prog.\ Phys.\  {\bf 65}, 1025 (2002)
  [arXiv:astro-ph/0204527].

\bibitem{Waxman:1998yy}
  E.~Waxman and J.~N.~Bahcall,
  Phys.\ Rev.\ D {\bf 59}, 023002 (1999)
  [arXiv:hep-ph/9807282].

\bibitem{Ahlers:2005sn}
  M.~Ahlers, L.~A.~Anchordoqui, H.~Goldberg, F.~Halzen, A.~Ringwald
  and T.~J.~Weiler,
  Phys.\ Rev.\ D {\bf 72}, 023001 (2005)
  [arXiv:astro-ph/0503229].



\bibitem{Berezinsky:2002nc}
  V.~Berezinsky, A.~Z.~Gazizov and S.~I.~Grigorieva,
  Phys.\ Rev.\ D {\bf 74}, 043005 (2006)
  [arXiv:hep-ph/0204357];
  R.~U.~Abbasi {\it et al.}  [High Resolution Fly's Eye Collaboration],
  Phys.\ Rev.\ Lett.\  {\bf 92}, 151101 (2004)
  [arXiv:astro-ph/0208243];
  V.~Berezinsky, A.~Z.~Gazizov and S.~I.~Grigorieva,
  Phys.\ Lett.\ B {\bf 612}, 147 (2005)
  [arXiv:astro-ph/0502550].

\bibitem{Stecker:1991vm}
  F.~W.~Stecker, C.~Done, M.~H.~Salamon and P.~Sommers,
  Phys.\ Rev.\ Lett.\  {\bf 66}, 2697 (1991)
  [Erratum-ibid.\  {\bf 69}, 2738 (1992)];
  F.~W.~Stecker,
  Phys.\ Rev.\ D {\bf 72}, 107301 (2005)
  [arXiv:astro-ph/0510537].


\bibitem{Anchordoqui:2005is}
  L.~Anchordoqui and F.~Halzen,
  Annals  Phys. \ {\bf 321}, 2660 (2006)  
  [arXiv:hep-ph/0510389].


\bibitem{Halzen:2006mq}
  F.~Halzen,
  arXiv:astro-ph/0602132.


\bibitem{note} For ice, the \v Cerenkov light generated by the shower 
particles spreads over a volume of radius 130~m at $10^{4}~{\rm GeV}$ and 
460~m at $10^{10}~{\rm GeV}$ (i.e., the bubble radius grows by about 50~m per 
decade of energy).  Therefore, a contained direct hit by a neutrino 
with energy $\sim 10^{10}~{\rm GeV}$ will not saturate the km$^3$ detector 
volume. For details the reader is referred to Ref.~\cite{Halzen:2006mq}. 


\bibitem{Learned:1994wg} Recall that cosmic neutrino beams, generally
  expected to be $\nu_\mu$'s and $\nu_e$'s from the decays of pions,
  kaons and perhaps heavy flavours, will be democratically distributed
  among flavors by the time they reach Earth.  J.~G.~Learned and
  S.~Pakvasa,
  Astropart.\ Phys.\  {\bf 3}, 267 (1995)
  [arXiv:hep-ph/9405296].


\bibitem{Feldman:1997qc}
  G.~J.~Feldman and R.~D.~Cousins,
  Phys.\ Rev.\ D {\bf 57}, 3873 (1998)
  [arXiv:physics/9711021].




\bibitem{Preskill:1991tb}
J.~Preskill, P.~Schwarz, A.~D.~Shapere, S.~Trivedi and F.~Wilczek,
Mod.\ Phys.\ Lett.\ A {\bf 6}, 2353 (1991).


\bibitem{Amati:1999fv}
D.~Amati and J.~G.~Russo,
Phys.\ Lett.\ B {\bf 454}, 207 (1999)
[arXiv:hep-th/9901092].
Equivalence in the evaporation properties of highly excited strings and BHs has also been recently considered by
G.~Domokos and S.~Kovesi-Domokos,
arXiv:hep-ph/0307099.


\bibitem{Cornet:2001gy}
F.~Cornet, J.~I.~Illana and M.~Masip,
Phys.\ Rev.\ Lett.\  {\bf 86}, 4235 (2001).
[arXiv:hep-ph/0102065].



\bibitem{Anchordoqui:2003ur}
  L.~A.~Anchordoqui, J.~L.~Feng, H.~Goldberg and A.~D.~Shapere,
  arXiv:hep-ph/0309082.


\bibitem{notetau} It is important to stress that a $\tau$ going
  through the detector at high energies without decaying will not
  deposit as much energy in the detector as a comparable-energy muon,
  due to the mass difference. (The direct-pair production process
  scales inversely with the mass, so it dominates tau-lepton energy
  loss resulting in $1/20^{\rm th}$ the light produced by a muon).
  Because of this, a hadronic $\nu_\tau$-vertex followed by a $\tau$
  track which leaves the detector may be indistinguishable from a low energy
  muon track from Hawking radiation, and consequently not flagged as a 
  BH event. However, this potential background can be
  eliminated by taking advantage of Earth-skimming neutrinos and the
  complete isotropy in the $\nu_\tau$-channel.



\bibitem{Duff:1994jr}
  M.~J.~Duff and J.~Rahmfeld,
  Phys.\ Lett.\ B {\bf 345}, 441 (1995)
  [arXiv:hep-th/9406105];
  C.~G.~.~Callan, J.~M.~Maldacena and A.~W.~Peet,
  Nucl.\ Phys.\ B {\bf 475}, 645 (1996)
  [arXiv:hep-th/9510134];
  A.~Strominger and C.~Vafa,
  Phys.\ Lett.\ B {\bf 379}, 99 (1996)
  [arXiv:hep-th/9601029];
  C.~G.~.~Callan and J.~M.~Maldacena,
  Nucl.\ Phys.\ B {\bf 472}, 591 (1996)
  [arXiv:hep-th/9602043];
  J.~M.~Maldacena and A.~Strominger,
  Phys.\ Rev.\ Lett.\  {\bf 77}, 428 (1996)
  [arXiv:hep-th/9603060];
  J.~M.~Maldacena and A.~Strominger,
  Phys.\ Rev.\ D {\bf 55}, 861 (1997)
  [arXiv:hep-th/9609026].


\bibitem{Chamblin:2004zg} 
  A.~Chamblin, F.~Cooper and G.~C.~Nayak,
  Phys.\ Rev.\ D {\bf 70}, 075018 (2004)
  [arXiv:hep-ph/0405054].


\bibitem{Cheung:2001ue}
  K.~m.~Cheung,
  Phys.\ Rev.\ Lett.\  {\bf 88}, 221602 (2002)
  [arXiv:hep-ph/0110163];
  K.~Cheung,
  Phys.\ Rev.\ D {\bf 66}, 036007 (2002)
  [arXiv:hep-ph/0205033].





\bibitem{Harris:2003db}
  C.~M.~Harris, P.~Richardson and B.~R.~Webber,
  JHEP {\bf 0308}, 033 (2003)
  [arXiv:hep-ph/0307305].


\bibitem{Cavaglia:2006uk}
  M.~Cavaglia, R.~Godang, L.~Cremaldi and D.~Summers,
  arXiv:hep-ph/0609001.

\bibitem{Cardoso:2004zi}
  V.~Cardoso, M.~C.~Espirito Santo, M.~Paulos, M.~Pimenta and B.~Tome,
  Astropart.\ Phys.\  {\bf 22}, 399 (2005)
  [arXiv:hep-ph/0405056].




\bibitem{Halzen:2003fi}
  F.~Halzen and D.~Hooper,
  JCAP {\bf 0401}, 002 (2004)
  [arXiv:astro-ph/0310152].






\end{thebibliography}
\end{document}